\documentclass{article}
\usepackage{piner}
\usepackage[figuresright]{rotating}
\begin{document}

\submitted{To be published in The Astrophysical Journal, v537, Jul 1, 2000}
\title{Space VLBI Observations of 3C 279 at 1.6 and 5 GHz}

\author{B. G. Piner\altaffilmark{1}, P. G. Edwards\altaffilmark{2},
A. E. Wehrle\altaffilmark{1}, H. Hirabayashi\altaffilmark{2},
J. E. J. Lovell\altaffilmark{3}, \\ \& S. C. Unwin\altaffilmark{1}}

\altaffiltext{1}{Jet Propulsion Laboratory, California Institute of Technology, 4800 Oak
Grove Dr., Pasadena, CA 91109; B.G.Piner@jpl.nasa.gov, Ann.E.Wehrle@jpl.nasa.gov,
Stephen.C.Unwin@jpl.nasa.gov}

\altaffiltext{2}{Institute of Space and Astronautical Science, Sagamihara, Kanagawa 229-8510, Japan;
pge@vsop.isas.ac.jp, hirax@vsop.isas.ac.jp}

\altaffiltext{3}{Australia Telescope National Facility, PO Box 76, Epping NSW 1710, Australia;
Jim.Lovell@atnf.csiro.au}

\begin{abstract}
We present VLBI Space Observatory Programme (VSOP) observations of the gamma-ray blazar
3C~279 at 1.6 and 5~GHz made on 1998 January 9-10 with the HALCA satellite and ground
arrays including the Very Long Baseline Array (VLBA).
The combination of the VSOP and VLBA-only images at these two frequencies maps the jet structure
on scales from 1 to 100~mas.  On small angular scales the structure is dominated by
the quasar core and the bright secondary component `C4' located 3 milliarcseconds 
from the core (at this epoch) at a position angle of $-$115$^\circ$.
On larger angular scales the structure is dominated by a jet extending to the southwest,
which at the largest scale seen in these 
images connects with the smallest scale structure seen in VLA images.
We have exploited two of the main strengths of VSOP: the ability to obtain matched-resolution
images to ground-based images at higher frequencies and the ability to measure
high brightness temperatures.  A spectral index map was made
by combining the VSOP 1.6~GHz image with a matched-resolution VLBA-only
image at 5~GHz from our VSOP observation on the following day.
The spectral index map shows the core to have a highly inverted spectrum, with some areas
having a spectral index approaching the limiting value for synchrotron self-absorbed radiation of 
$\alpha = +2.5$ (where $S \propto \nu^{+\alpha}$).
Gaussian model fits to the VSOP visibilities revealed high brightness temperatures ($>10^{12}$~K) that 
are difficult to measure with ground-only arrays.  An extensive error analysis was performed
on the brightness temperature measurements.  Most components did not have measurable
brightness temperature upper limits, but lower limits were measured as high as $5~\times~10^{12}$~K.
This lower limit is significantly above both the nominal inverse
Compton and equipartition brightness temperature limits.  The derived Doppler factor, Lorentz factor, and
angle to the line-of-sight in the case of the equipartition limit
are at the upper end of the range of expected values for EGRET blazars. 
\end{abstract}

\keywords{quasars: individual (3C~279) --- galaxies: active ---
galaxies: jets --- techniques: interferometric --- radio continuum: galaxies ---
radiation mechanisms: non-thermal}

\section{Introduction}
\label{intro}
The quasar 3C~279 ($z$=0.536) is one of the most intensively studied quasars for several reasons.
It was the first radio source 
observed to exhibit
superluminal motion (Knight et al.\ 1971; Whitney et al.\ 1971; Cohen et al.\ 1971).  It was
the first blazar --- and remains one of the brightest --- detected in high-energy 
$\gamma$-rays by the EGRET instrument on the {\em Compton Gamma Ray Observatory}
(Hartman et al.\ 1992).  The EGRET detection prompted several large multiwavelength studies
of this source.  Results of these studies are presented by Maraschi et al.\ (1994),
Grandi et al.\ (1996), Hartman et al.\ (1996), and Wehrle et al.\ (1998).  
The radio flux density of 3C~279 has been monitored by the Michigan group since 1965
at frequencies of 4.8, 8.0, and 14.5 GHz (Aller et al.\ 1985).
The observations presented in this paper occurred at the beginning of a total flux density flare
that would later reach the highest flux densities yet recorded in this program for 3C~279.
\footnote{http://www.astro.lsa.umich.edu/obs/radiotel/gif/1253-055.gif}

Following the discovery of superluminal motion,
the parsec scale structure of 3C~279 has been monitored 
using the VLBI technique.
Cotton et al.\ (1979)
measured a speed of 15~$c$ for the original superluminal jet component. 
(Throughout the paper we assume $H_{0}$=70 km s$^{-1}$ Mpc$^{-1}$ and $q_{0}$=0.1,
and component speeds measured by others have been expressed in these terms.)
Unwin et al.\ (1989) and Carrara et al.\ (1993) 
describe VLBI monitoring of 3C~279 at 5, 11, and 22~GHz throughout the 1980s.
These authors observed the motions of several new superluminal components, and
found that the speeds of these components were only one-quarter to one-third (3-5~$c$)
of that measured for the original superluminal component during the 1970s.  
VLBI monitoring of 3C~279 has
been undertaken at 22 and 43~GHz during the 1990s.  Initial results of this high-frequency
monitoring are reported by Wehrle, Unwin, \& Zook (1994) and Wehrle et al.\ (1996), and
final results will be reported by Wehrle et al.\ (2000).  
High-resolution VLBI polarimetric images of 3C~279 have been made by
Lepp\"{a}nen, Zensus, \& Diamond (1995), Lister, Marscher, \& Gear (1998)
and Wardle et al.\ (1998).  
The detection of circularly polarized radio emission by Wardle et al.\ provides
some of the first direct evidence that electron-positron pairs are an
important component of the jet plasma.
One notable feature of the VLBI 
observations of this source has been the differing speeds and position angles of the
VLBI components.  Carrara et al.\ (1993) and Abraham \& Carrara (1998) claim that these
can be explained by ejection of components by a precessing jet.  

3C~279 was observed during the TDRSS space VLBI experiments
at 2.3\,GHz (Linfield et al.\ 1989) and 15 GHz (Linfield et al.\ 1990), 
with source frame brightness temperatures between
1.6 and 2.0~$\times$~10$^{12}$~K being measured for this source.
A brightness temperature of 1.9~$\times$~10$^{12}$~K 
(translated from observed frame to source frame) was also measured for 3C~279 in the 
22~GHz VSOP Pre-Launch Survey (Moellenbrock et al.\ 1996).
This source has been detected in VLBI observations up to frequencies of 215~GHz (Krichbaum et al.\ 1997).
This paper reports on the VSOP observations of 3C~279 made
during the first Announcement of Opportunity period (AO1).
Hirabayashi et al.\ (1999) and Edwards et al.\ (1999) presented preliminary analyses of these
5 and 1.6 GHz VSOP observations respectively.  Here we present higher dynamic range images
together with analysis and interpretation of model fits and a spectral index map.

\section{Observations}
\label{obs}

The quasar 3C~279 was observed on two consecutive days during the AO1 phase of the VSOP
mission: on 1998 January 9 at 1.6~GHz, and on 1998 January 10 at 5~GHz.  The VSOP mission 
uses the Japanese HALCA satellite as an element in a 
changeable VLBI array in order to obtain visibility measurements
on baselines larger than the Earth's diameter.  HALCA was launched on 1997 February 12 and
carries an 8 meter antenna through an elliptical orbit
with an apogee height of 21,400~km (yielding baselines up to 2.6 Earth diameters),
a perigee height of 560~km, and an orbital
period of 6.3 hours.  HALCA has operational observing bands at 1.6 and 5~GHz (18 and 6 cm).
The data from the satellite are recorded by a network of ground tracking
stations and subsequently correlated with the data from the participating ground telescopes.
The VSOP system and initial science results are discussed by Hirabayashi et al.\ (1998).

The observations of 1998 January 9--10 were conducted using the standard VSOP observing mode:
two 16 MHz intermediate frequency bands, each 2-bit sampled at the 
Nyquist rate in left circular polarization, for a total
data rate of 128 Mbps.  The ground telescope arrays were made up of nine elements of the 
NRAO Very Long Baseline Array\footnote{The National Radio Astronomy Observatory
is a facility of the National Science Foundation operated
under cooperative agreement by Associated Universities, Inc.} (VLBA) --- Hancock did not observe
because of a power failure --- with the addition of the 70 m telescopes at Goldstone, California, U.S.A.
and Tidbinbilla, Australia on January 9, and the 64 m telescope at Usuda, Japan on January 10.
The total observing time on January 9
was 6 hours, including 3.5 hours of HALCA data from the tracking station at Green Bank, West Virginia,
covering the portion of HALCA's orbit from near perigee to near apogee.  
The data from Goldstone were not used in the final image as it observed for only
a short time and there were significant calibration uncertainties with this data.
The total 
observing time on January 10 was 10 hours, although for the final 1.5 hours only Mauna Kea, Usuda and
HALCA observed and this data was also excluded from the final image (as four telescopes are
required for amplitude self-calibration).
This observation included two HALCA tracking passes
of 3 hours each (separated by 4 hours) by the tracking stations at Robledo, Spain and
Goldstone, California, U.S.A.  
Each tracking pass covered the portion of HALCA's orbit from near perigee to near apogee.
The data from both observations were correlated at the VLBA correlator in Socorro.
Through the remainder of this paper, we use ``VSOP'' to refer to the HALCA+ground-array combination.

Calibration and fringe-fitting were done with
the AIPS software package.  3C~279 is a strong source, and good fringes
were found to HALCA during all tracking passes.  The antenna gain
for Saint Croix had to be adjusted upward by a factor of $\sim$~2 from its nominal value at 
5~GHz since the VSOP observations were conducted at 4800--4832\,MHz because HALCA's performance
is better at these frequencies.
Figure~1 shows the $(u,v)$ plane coverages
of the two observations.  These $(u,v)$ plane coverages result in highly elliptical beams:
the major to minor axis ratio of the 5~GHz beam is 8:1.  The addition of Tidbinbilla to the 
1.6~GHz observation improves the north-south coverage and reduces this ratio to 4:1.
Plots of correlated flux density vs. $(u,v)$ distance projected along a position angle of $-115\arcdeg$
(the position angle of the brightest structure) are shown in Figure~2.  The addition of HALCA
extends the projected $(u,v)$ distances compared to the ground-only baselines
from 120 to 340 M$\lambda$ at 5~GHz, and from
65 to 120 M$\lambda$ at 1.6~GHz.  The beating evident in the correlated flux densities
indicates that on milliarcsecond scales the morphology is dominated by two components
of similar flux density.

\begin{figure*}[!t]
\plotfiddle{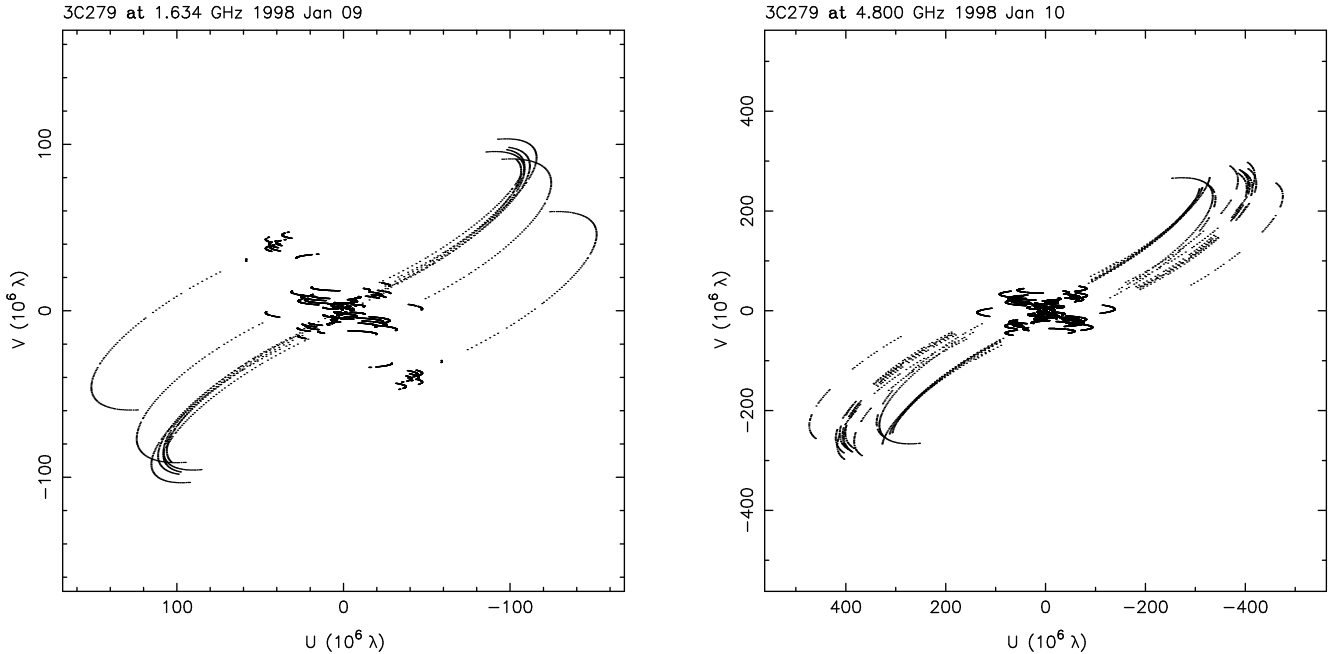}{3.5 in}{-90}{70}{70}{-263}{350}
\caption{$(u,v)$ plane coverages for the VSOP observations of 3C~279.  The left
panel shows the 1.6~GHz coverage, the right panel the 5~GHz coverage.}
\end{figure*}

\begin{figure*}[!t]
\plotfiddle{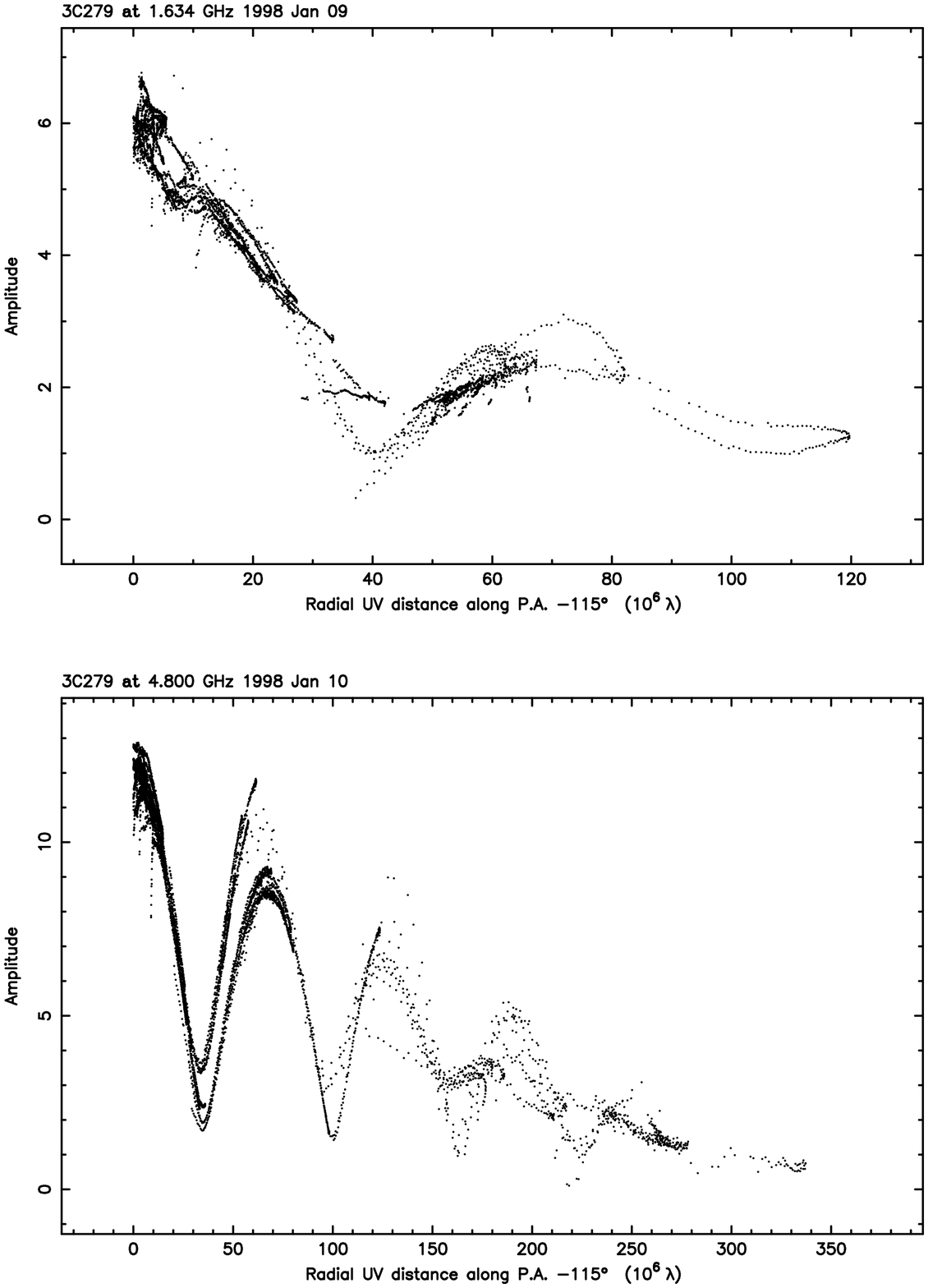}{5.0 in}{0}{50}{50}{-140}{-9}
\caption{Correlated flux density in Janskys vs. projected $(u,v)$ distance along position
angle $-115\arcdeg$.  The top panel shows the 1.6~GHz observation, the bottom the
5~GHz observation.}
\end{figure*}

\section{Results}
\label{results}
\subsection{Images}
\label{images}
Images from these datasets were produced using standard CLEAN and
self-calibration procedures from the Difmap software package (Shepherd, Pearson, \& Taylor 1994).
Figure~3 shows two images from the 5~GHz observation on 1998 January 10: the 
full-resolution space VLBI image and the VLBA-only image made from the same
dataset with the space baselines removed.  Figure~4 shows the full-resolution space VLBI and
the VLBA-only images from the 1.6~GHz observation on the previous day
(Tidbinbilla baselines were also removed from this ground-only image to reduce
the size of the $(u,v)$ holes).  
The VSOP images are displayed using uniform weighting (although cleaning was done with both
uniform and natural weighting to model the extended structure), while the VLBA images are
displayed with natural weighting.  We stress the importance of using uniform weighting
with VSOP datasets.  A naturally weighted
VSOP image degrades the resolution of the ground-space array because of the
higher weighting of the larger ground antennas and the much denser sampling of the $(u,v)$ plane
on the ground baselines.

\begin{figure*}[!t]
\plotfiddle{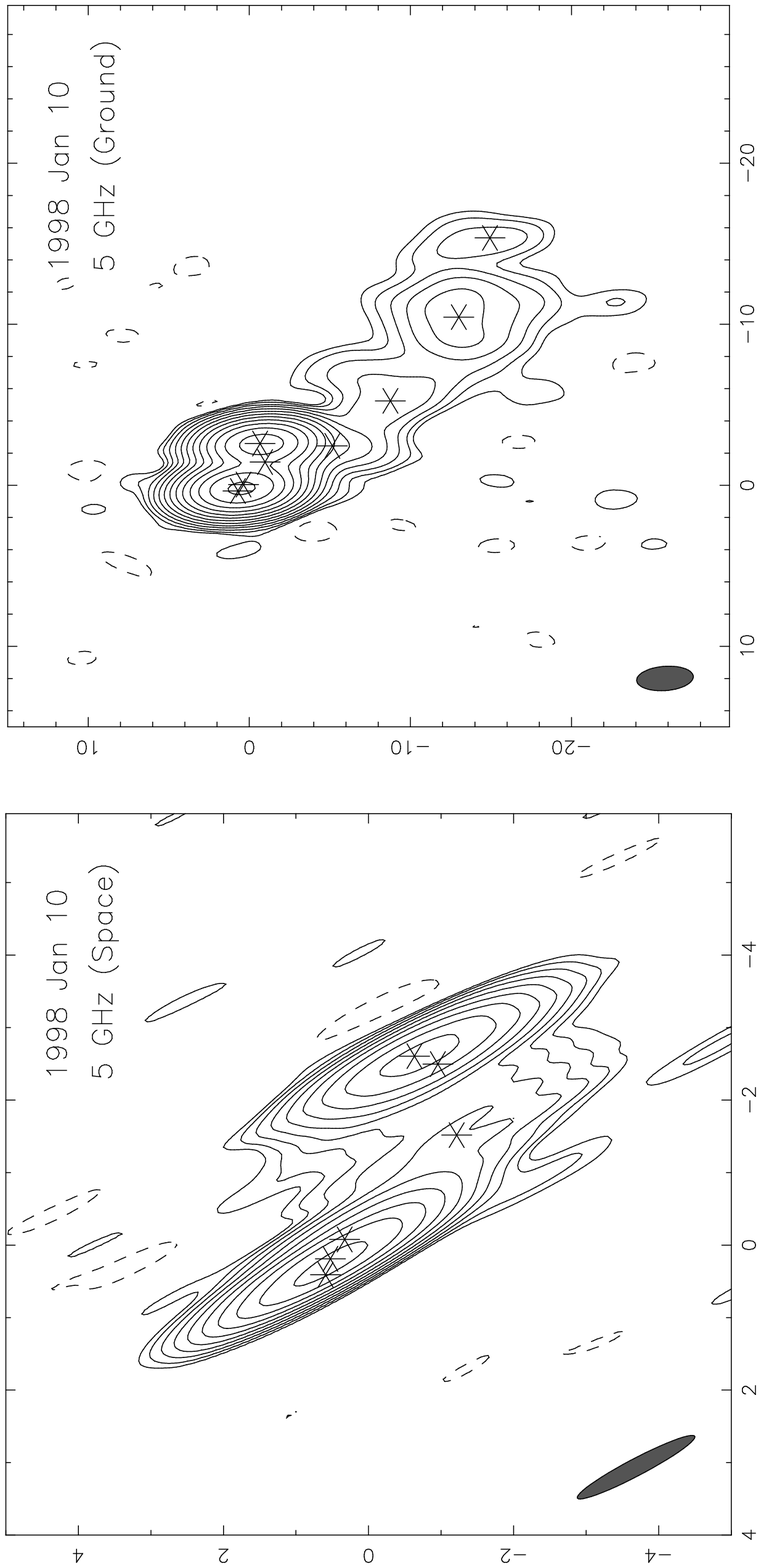}{3.125 in}{-90}{65}{65}{-263}{305}
\caption{Images of 3C~279 from the 5~GHz VSOP observation on 1998 January 10.  The image on
the left is the full-resolution space VLBI image, the image on the right is the VLBA-only image
made from the same dataset with the space baselines removed.  The lowest contour in each image has
been set equal to 3 times the rms noise level in that image.  The peak flux densities are 3.98 and 6.67
Jy beam$^{-1}$, the contour levels are 6.0 mJy beam$^{-1}$
$\times$ 1,2,4,...512 and 1.4 mJy beam$^{-1}$ $\times$ 1,2,4,...4096,
and the beam sizes are 1.83~$\times$~0.24~mas at 28$\arcdeg$ and 3.52~$\times$~1.51~mas
at 4$\arcdeg$ for the space and ground images respectively.
Model-fit component positions are marked with asterisks.}
\end{figure*}

\begin{figure*}
\plotfiddle{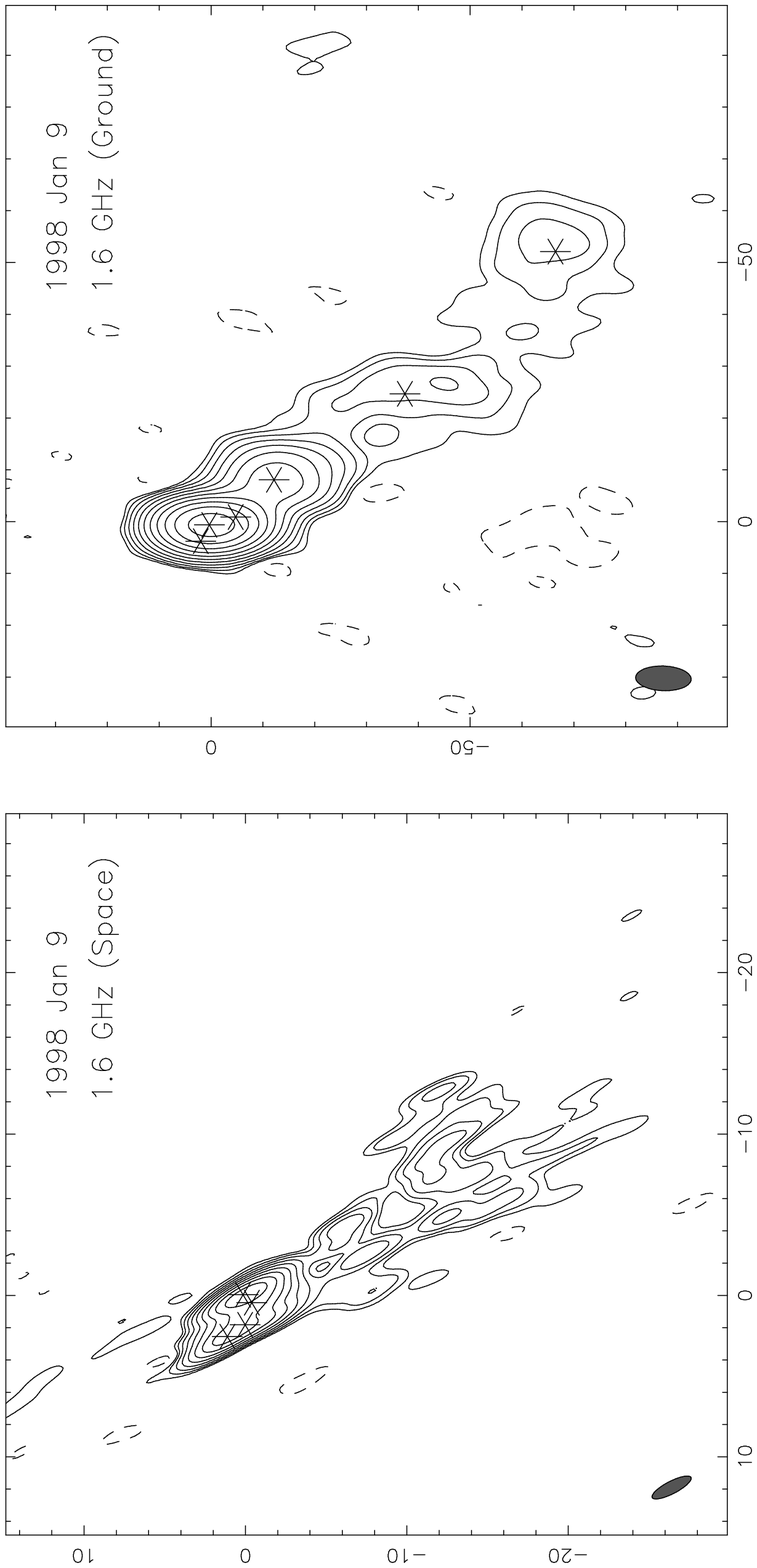}{3.125 in}{-90}{65}{65}{-263}{305}
\caption{Images of 3C~279 from the 1.6~GHz VSOP observation on 1998 January 9.  The image on
the left is the full-resolution space VLBI image, the image on the right is the VLBA-only image
made from the same dataset with the space
and Tidbinbilla baselines removed.  The lowest contour in each image has
been set equal to 3 times the rms noise level in that image.  The peak flux densities are 2.29 and 4.73
Jy beam$^{-1}$, the contour levels are 4.7
mJy beam$^{-1}$ $\times$ 1,2,4,...256 and 2.9 mJy beam$^{-1}$ $\times$ 1,2,4,...1024,
and the beam sizes are 2.71~$\times$~0.77~mas at 28$\arcdeg$ and 10.7~$\times$~4.77~mas
at $-2\arcdeg$ for the space and ground images respectively.
Model-fit component positions are marked with asterisks.}
\end{figure*}

The VSOP and VLBA-only images differ in scale by about a factor of four,
showing the greatly increased resolution provided by the space baselines.  
The VSOP images presented in Figures~3 and 4 are the highest resolution images yet produced
of 3C~279 at these frequencies.  The 1.6 and 5~GHz VSOP images have formal dynamic ranges (peak:rms) of
1,500:1 and 2,000:1 respectively, demonstrating that high dynamic range images can be made 
despite the small size of the HALCA orbiting antenna.  For comparison, the VLBA-only images have dynamic
ranges of 5,000:1 and 15,000:1 respectively.  The 5~GHz VSOP image
shows that on small angular scales 3C~279 is dominated
by a double structure.  This structure consists of the compact
core and inner jet region (the feature to the east) and a bright jet component about 3~mas from
the core along a position angle of $-115\arcdeg$ (the feature to the west).  
This bright jet component, `C4', is a well known feature first observed
in 1985.  This component will be discussed further in $\S$~\ref{comps}.

The structure on slightly larger angular scales is quite different.
The 5~GHz VLBA image and the 1.6~GHz VSOP image both show the double structure mentioned above
(although in the 1.6~GHz image the jet component is brighter than the core which results in the jet component
being placed at the phase center),
as well as a more extended jet to the southwest along a position angle of approximately 
$-140\arcdeg$.  This position angle is similar to that seen in older VLBI images (see $\S$~\ref{comps}) as well as 
VLA and MERLIN images (de Pater \& Perley 1983; Pilbratt, Booth, \& Porcas 1987; Akujor et al.\ 1994).  
The resolutions of the 5~GHz VLBA image
and the 1.6~GHz VSOP image are roughly equal, a fact that allows a spectral index map to be made from
these images ($\S$~\ref{index}).  The jet emission in the 1.6~GHz VSOP image is quite complex,
and it appears in Figure~4 that the jet may be limb brightened.  
However, we caution against over-interpreting these
complex jet features because the CLEAN striping produced by the holes in the $(u,v)$ plane coverage
runs parallel to the jet, and because simulations indicate that,
due to the lack of complete $(u,v)$ plane coverage,
space VLBI images may show such knotty structure
when the actual brightness distribution is smoother (D. Murphy 1999, private communication).
The 1.6~GHz VLBA image shows structure extending out to $\sim$~100~mas from the core, all the way
out to the smallest size scales sampled by the 22\,GHz VLA images of de Pater \& Perley (1983).

\subsection{Model Fits}
\label{mfit}
The Difmap model-fitting routine was used to fit elliptical Gaussian components to the visibility data
for each image in Figures~3 and 4.  When model fitting VSOP data, it is important to increase the weight
of HALCA over the default weighting used by Difmap in order for the space baseline data to
have any effect on the model.
We increased the weight of HALCA during model fitting by a factor
equal to the product of the ratios of the average ground baseline sensitivity to the average space baseline
sensitivity ($\sim$~50 for these observations)
and the number of ground visibilities to the number of space visibilities 
($\sim$~10 for these observations).  This effectively achieves a ``uniform'' weighting during
model fitting, and causes the space visibilities to have an effect on the model fitting
equal to that of the ground visibilities.

The results of the model fitting are given in Table~\ref{mfittab}.  Component numbers are given
only for ease of later reference, and are not meant to identify the same component between images.
Tentative component identifications obtained from the discussion in $\S$~\ref{comps} are
given in the third column.  Note that the lower resolution images (e.g. the 1.6 GHz VLBA image)
may not properly split the flux between the core and the closest component.
In each case we have taken the far northeastern component to be the core, and have defined the other
component positions relative to the position of the core.  In the 1.6~GHz VSOP image,
the southwestern jet is too complex to be fit with simple Gaussian components, and we left
the CLEAN components in this region during model fitting.  When fitting elliptical components,
the model fitting chi-squared statistic
is frequently minimized by an ellipse of zero axial ratio.  This is unphysical in
the sense that these components have formally infinite brightness temperatures.  In these cases
an upper limit to the size of the component can be used instead of the best-fit value to find a 
lower limit to the brightness temperature.  Since we use an error analysis method to find these
limits in $\S$~\ref{tb}, we have left these zero axial ratio components in the models if
they minimize the chi-squared for that model.  Table~\ref{mfittab} also gives the source frame
brightness temperatures for the VSOP models, where the maximum brightness temperature of
a Gaussian component is given by
\begin{equation}
\label{tbeq}
T_{B}=1.22\times10^{12}\;\frac{S(1+z)}{ab\nu^{2}}\;\rm{K},
\end{equation}
where $S$ is the flux density of the component in Janskys,
$a$ and $b$ are the FWHMs of the major and minor axes respectively in mas,
$\nu$ is the observation frequency in GHz, and $z$ is the redshift.

\begin{table*}[!t]
\caption{Gaussian Models}
\label{mfittab}
\begin{center}
\begin{tabular}{l c l r r r r r r r} \tableline \tableline
& & \multicolumn{1}{c}{Tentative\tablenotemark{a}} 
& \multicolumn{1}{c}{$S$\tablenotemark{b}} & \multicolumn{1}{c}{$r$\tablenotemark{c}}
& \multicolumn{1}{c}{PA\tablenotemark{c}}
& \multicolumn{1}{c}{$a$\tablenotemark{d}}
& & \multicolumn{1}{c}{$\Phi$\tablenotemark{e}} & \multicolumn{1}{c}{$T_{B}$\tablenotemark{f}} \\ 
Image & Comp. \# & \multicolumn{1}{c}{ID} 
& \multicolumn{1}{c}{(Jy)} & \multicolumn{1}{c}{(mas)}
& \multicolumn{1}{c}{(deg)}
& \multicolumn{1}{c}{(mas)} & \multicolumn{1}{c}{$b/a$} & \multicolumn{1}{c}{(deg)} &
\multicolumn{1}{c}{($10^{12}$ K)} \\ \tableline
5 GHz VSOP        & 1 & Core & 1.71 &  0.00 &    ...  &  0.28 & 0.00 &   29.8 &  ... \\ 
                  & 2 &      & 3.91 &  0.23 & --107.1 &  0.56 & 0.35 &   38.6 & 2.90 \\ 
	          & 3 &      & 1.36 &  0.55 & --118.4 &  0.83 & 0.35 &   44.4 & 0.46 \\ 
	          & 4 & C3?  & 0.83 &  2.64 & --133.2 &  2.53 & 0.52 &   36.4 & 0.02 \\ 
	          & 5 & C4   & 3.39 &  3.26 & --112.1 &  0.46 & 0.50 & --12.7 & 2.63 \\ 
	          & 6 & C4   & 0.79 &  3.29 & --118.1 &  0.97 & 0.25 &   34.3 & 0.28 \\ 
5 GHz VLBA        & 1 & Core & 4.17 &  0.00 &    ...  &  0.29 & 0.00 &   41.5 &      \\ 
	          & 2 &      & 2.87 &  0.53 & --130.9 &  0.38 & 0.00 & --81.7 &      \\ 
	          & 3 & C3?  & 0.64 &  2.47 & --133.1 &  1.50 & 0.71 &    2.0 &      \\ 
	          & 4 & C4   & 4.23 &  3.25 & --115.0 &  0.63 & 0.51 &  --2.3 &      \\ 
	          & 5 & C2?  & 0.06 &  6.50 & --154.4 &  2.78 & 0.33 & --12.1 &      \\ 
	          & 6 & C2?  & 0.05 & 11.00 & --149.4 &  3.47 & 0.56 &   34.0 &      \\ 
	          & 7 & C1?  & 0.14 & 17.45 & --141.8 &  4.58 & 0.85 &  --7.1 &      \\ 
	          & 8 &      & 0.01 & 22.18 & --134.9 &  3.15 & 0.29 &    6.4 &      \\
1.6 GHz VSOP
\tablenotemark{g} & 1 & Core & 0.76 &  0.00 &    ...  &  1.95 & 0.00 &   34.4 &  ... \\ 
	          & 2 &      & 0.92 &  1.33 & --146.4 &  2.29 & 0.35 &   20.5 & 0.35 \\ 
	          & 3 & C3?  & 1.57 &  2.59 & --125.9 &  1.93 & 0.16 &   55.4 & 1.92 \\ 
	          & 4 & C4   & 2.21 &  2.77 & --110.6 &  0.95 & 0.00 &   43.1 &  ... \\
1.6 GHz VLBA      & 1 & Core & 0.23 &  0.00 &    ...  & 10.64 & 0.00 &    5.9 &      \\ 
	          & 2 & C4   & 5.09 &  3.55 & --117.9 &  2.71 & 0.51 &   55.8 &      \\ 
	          & 3 & C2?  & 0.35 &  8.20 & --145.3 &  7.69 & 0.00 &   25.0 &      \\ 
	          & 4 & C1?  & 0.61 & 18.45 & --140.0 &  7.44 & 0.45 &   64.6 &      \\ 
	          & 5 &      & 0.13 & 48.58 & --144.2 & 27.19 & 0.46 &   33.2 &      \\ 
	          & 6 & `C'  & 0.14 & 88.35 & --140.8 & 60.26 & 0.28 &   65.1 &      \\ \tableline
\end{tabular}
\end{center}
\tablenotetext{a}{Tentative component identifications from the discussion in $\S$~\ref{comps}.
A question mark indicates a more speculative identification.}
\tablenotetext{b}{Flux density in Janskys.  Note that the lower resolution images (e.g. the 1.6 GHz VLBA image)
may not properly split the flux between the core and the closest component.}
\tablenotetext{c}{$r$ and PA are the polar coordinates of the
center of the component relative to the presumed core.
Position angle is measured from north through east.}
\tablenotetext{d}{$a$ and $b$ are the FWHM of the major and minor axes of the Gaussian
component.}
\tablenotetext{e}{Position angle of the major axis measured from north through east.}
\tablenotetext{f}{Maximum source frame Gaussian brightness temperatures are given
for the VSOP models.}
\tablenotetext{g}{The southwestern jet is fit by clean components in this model.}
\end{table*}

\section{Discussion}
\label{discussion}
\subsection{Identification of Historical Components}
\label{comps}
In this section we consider the identification of components seen in these images
with previously published VLBI components, using these components' published positions and
velocities.  We work from the innermost components outward, starting with the components
fit to the 5~GHz VSOP image.  We caution that any such identifications are highly
speculative, particularly for the older components where much time has elapsed since the
last published image.  Note that prior to completion of the NRAO VLBA (1995), global VLBI
network sessions at 22 GHz occurred only twice per year and used less than half a dozen antennas.

A total of six elliptical Gaussian components are required to fit the 5~GHz VSOP data.  The first three of these
components are interior to 1~mas, and represent the core and two components of the inner jet.
The region interior to 1~mas has been studied by Wehrle et al.\ (2000) at 22 and 43 GHz
and Rantakyr\"{o} et al.\ (1998) at 86 GHz, and
they find it to be a complex region with multiple components.  
Attempts to name components in this region have resulted in some confusion.  The component named
`C5' in three 11 GHz maps from 1989.3-1991.1 by Carrara et al.\ (1993) and Abraham \& Carrara (1998) 
is not the same component referred to
as either `C5' by Wehrle et al.\ (1994) at 22 GHz, 
Lepp\"{a}nen et al.\ (1995) at 22 GHz, and Lister et al.\ (1998) at 43 GHz or
`the stationary 1~mas feature' by Wehrle et al.\ (1996) at 22 GHz.  
Wehrle et al.\ (1994) and Wehrle et al.\ (1996)
identify a component between C5 and the core, and Lepp\"{a}nen et al.\ (1995) identify two components
in this region which they name `C6' and `C7'.  Lister et al.\ (1998) do not detect these components
but record another new component named `C8'.  Clarification of the components in this region 
must await completion of the Wehrle et al.\ (2000) analysis.

The situation beyond $\sim$~1~mas is easier to interpret.  The bright feature at
$\sim$~3~mas is the component C4 that has been seen by many other authors (Unwin et al.\ 1989; 
Carrara et al.\ 1993; Wehrle et al.\ 1994; Lepp\"{a}nen et al.\ 1995; Wehrle et al.\ 1996; Lister et al.\ 1998;
Wardle et al.\ 1998; Kellermann et al.\ 1998).
This component has been moving along a position angle of $-115\arcdeg$ for over 10 years.
The brightness distribution of C4 is asymmetric, with the leading edge being sharper than the trailing
edge, and two model components are required to represent it (components~5 and 6 in the 5~GHz VSOP model).
The 43~GHz data of Wehrle et al.\ (2000) also require two components to represent the structure of C4.
The sharp leading edge of this component is suggestive of a working surface or shock front.  The 
polarization observations of Lepp\"{a}nen et al.\ (1995), Lister et al.\ (1998), and Wardle et al.\ (1998)
show that C4 has
a magnetic field transverse to the jet, also indicative of a shock front.  
Carrara et al.\ (1993) determined a motion of
0.15$\pm$0.01~mas/yr (4~$c$) for C4, however the position of C4 in our 5\,GHz image
is inconsistent with a simple extrapolation at this speed.
However, we note that at the time of the observations of Carrara et al.\
C4 was $\sim$1\,mas from the core and in the region noted above as being
difficult to interpret.
More recent data from Wehrle et al. (2000) and Kellermann (1999, private communication)
indicate a speed of $\sim$7~$c$ for C4 between 1991 and 1999, which is consistent with
the position seen in the VSOP images in this paper.
Although the larger scale structure
has a much different position angle than C4, C4 shows no signs of altering its path
to follow the larger scale structure, and appears to be continuing along a position angle of $-115\arcdeg$.
Component C4 
dominates the emission in the 1.6~GHz VSOP image and
it is represented by component~4 in the 1.6~GHz VSOP model.

Older VLBI measurements (Cotton et al.\ 1979; Unwin et al.\ 1989; Carrara et al.\ 1993) followed a series of
components (C1--C3) moving along position angles of $-130$ to $-140\arcdeg$.  Since this is also the position
angle of the larger scale structure seen in our images, it is reasonable that this string of components
may form the southwestern jet in our images.  
The 5\,GHz VSOP image shows 
a fainter, more diffuse
component at a position angle of $-133\arcdeg$ located $\sim$2.6\,mas from the core
(component~4 in the 5\,GHz model, component~3 in the 1.6~GHz model).
Emission is also seen at this position in the images of 
Kellermann et al.\ (1998) and Wehrle et al.\ (2000).
An attempt to identify this component with the most recent component 
ejected along this position angle (C3) would
imply a drastic deceleration for C3; a straightforward extrapolation of the motion of C3 given by 
Carrara et al.\ (1993)
would place it at a separation of 4~mas in 1998 and so we consider such an identification unlikely.  
An extrapolation of the motion of C2 estimated
from the positions given by Unwin et al.\ (1989) would place C2 between 6 and 10~mas from the core at the
epoch of our VSOP observations, and conceivably identify it with either component~5 or 6 
in the 5~GHz VLBA model
(component~3 in the 1.6~GHz VLBA model).  Cotton et al.\ (1979) reported a large proper motion of
0.5~mas/yr during the early 1970s; their data was re-examined and their interpretation judged to be correct
by Unwin et al.\ (1989).  An extrapolation the motion of the Cotton et al.\ (1979) component (which could also
be the component C1 of Unwin et al.\ [1989] located $\sim$8\,mas from the core at 1984.25) 
would place it about 15~mas from the core in 1998, and mean that
it could be identified with the relatively bright component~7 of the 5~GHz VLBA model (component~4
in the 1.6~GHz VLBA model).  The uncertainties of extrapolating component motions 
makes this analysis highly speculative; the validity of the assumption of
constant projected speed with time is unclear, and enough time has elapsed 
that the fate of these older components will probably never be certain.

A series of VLA maps of 3C~279 at varying resolution has been published by de Pater \& Perley (1983).
Their highest resolution map shows a component (component `C') 95~mas from the core
in position angle $-145\arcdeg$.  (This map has a higher resolution than the MERLIN maps of
Pilbratt et al.\ [1987] and Akujor et al.\ [1994] because it is at a higher frequency).
Our 1.6~GHz VLBA image extends out to this distance from the core,
and we see a 140 mJy component at 88~mas in position angle $-141\arcdeg$  
(component~6 in the 1.6~GHz VLBA model).  We can tentatively identify this
with component C of de Pater \& Perley (1983), given the large errors implied by their VLA beam 
(60~mas resolution) and the similar size of our model-fit component. These errors
make it impossible to infer anything about the motion of this component between 1982 and 1998.
We have, however, matched the largest scale VLBI structures with the smallest scale VLA structures, 
and established a continuous connection between the parsec and kiloparsec scales in this source.

\subsection{Spectral Index Map}
\label{index}
Construction of spectral index maps is often hindered by the differing resolutions of the images
at different frequencies; however, a unique capability of the VSOP mission is that it can provide matched
resolution images to ground-based images at higher frequencies, enabling
the construction of a spectral index map from two images of approximately equal resolution.
We have used this capability to produce a spectral index map from the 1.6~GHz VSOP image and the
5~GHz VLBA-only image; this spectral index map is shown in Figure~5.  
These two datasets were taken only 1 day apart, so the errors
due to component motions are  negligible.  To produce the spectral index map
the two images were restored with their average beam of 
3.12~$\times$~1.14~mas with a major axis position angle of 15.9$\arcdeg$, 
each image was restored from the clean components without residuals (the flux left over after
the clean components convolved by the dirty beam have been subtracted from the data),
and no spectral index was calculated for pixels where the flux densities were less than 3 times the rms noise level
in the images at both frequencies.  
(A spectral index map using images restored from clean components {\em with\/} residuals
was also produced and gave essentially the same results. We prefer to use the images
without residuals as the residuals can then be used to assess the possible errors 
in the spectral index map.)
In the following discussion we use $S\propto\nu^{+\alpha}$.

\begin{figure*}[!t]
\plotfiddle{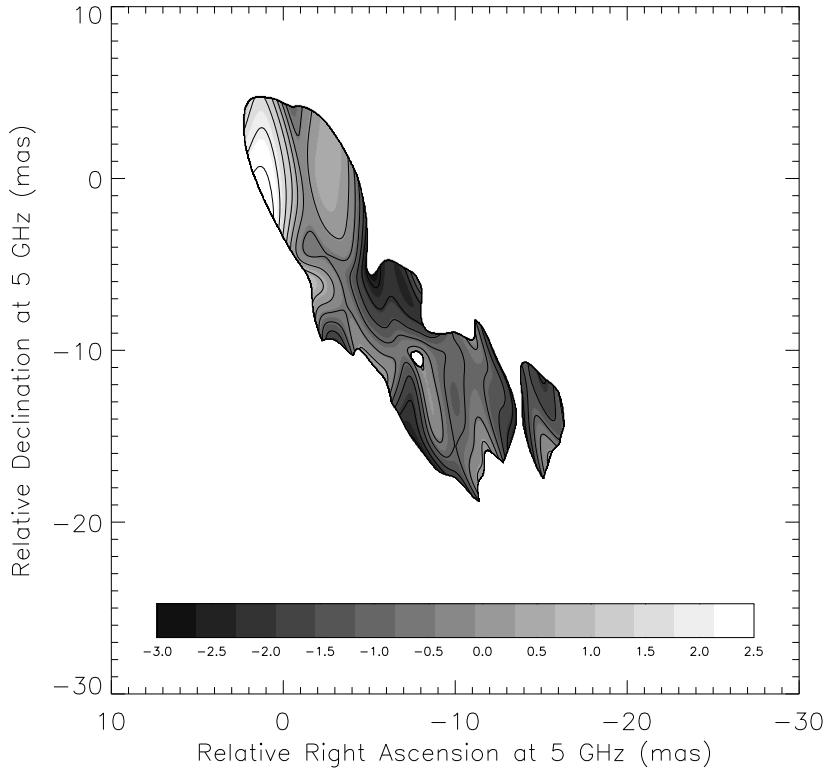}{4.0 in}{0}{60}{60}{-162}{-72}
\caption{Spectral index map of 3C~279 made from the 1.6~GHz VSOP image and the 5~GHz VLBA image.
The gray-scale color bar along the bottom of the image
indicates the value of the spectral index ($S\propto\nu^{+\alpha}$), with lighter colors
indicating an inverted spectrum and darker colors a steep spectrum.  Spectral index contours are also
plotted at intervals of 0.5, from $-2.5$ to 2.5.  The beam used was the average of the
1.6~GHz VSOP beam and the 5~GHz VLBA beam, or 3.12~$\times$~1.14~mas at 15.9$\arcdeg$.
The white pixels near the center of the jet represent an area where the flux was below the clipping
level applied for calculation of the spectral index.}
\end{figure*}

A major difficulty in making spectral index maps lies in correctly registering the two images.
We investigated several alignments of these two images, including aligning the peak core pixels
and aligning the peak pixels in the bright jet component (C4).  We doubled the number of pixels
across each image in order to measure the required shifts as accurately as possible.
Aligning the peak core pixels produced unphysical results, including a highly inverted spectral index
along the right edge of component C4.  Of the different alignments tried, aligning the peak pixels in the  
jet component C4 produced the most physically reasonable results.  
The reason for this can be seen {\em a posteriori\/} from
Figure~5.  The spectral index is constant across component C4, meaning that the brightest pixels in C4 at
1.6 and 5~GHz will represent the same physical location.  On the other hand, there are steep spectral
index gradients across the core region, so the peak pixel in the core region will be at different
locations at 1.6 and 5~GHz.  

We also constructed a map of the error in the spectral index; this error map is shown in Figure~6.
The error was calculated by standard propagation of errors, using the fluxes at each pixel
in each residual map as the flux errors.  This method actually gives a lower limit to the
error at each pixel, because it does not take into account calibration errors or errors in imaging the
source structure caused by the holes in the $(u,v)$ plane coverage.  In the region comprising the southwestern
jet the errors in the spectral index $\alpha$
range from $\pm 0.05$ in the brighter parts of the jet (the knot at $\sim$~18~mas)
to $\pm 0.3$ in the fainter parts.  The formal errors in the core and C4 regions are quite low
and so the errors in these regions will be dominated by the other effects mentioned above.

\begin{figure*}[!t]
\plotfiddle{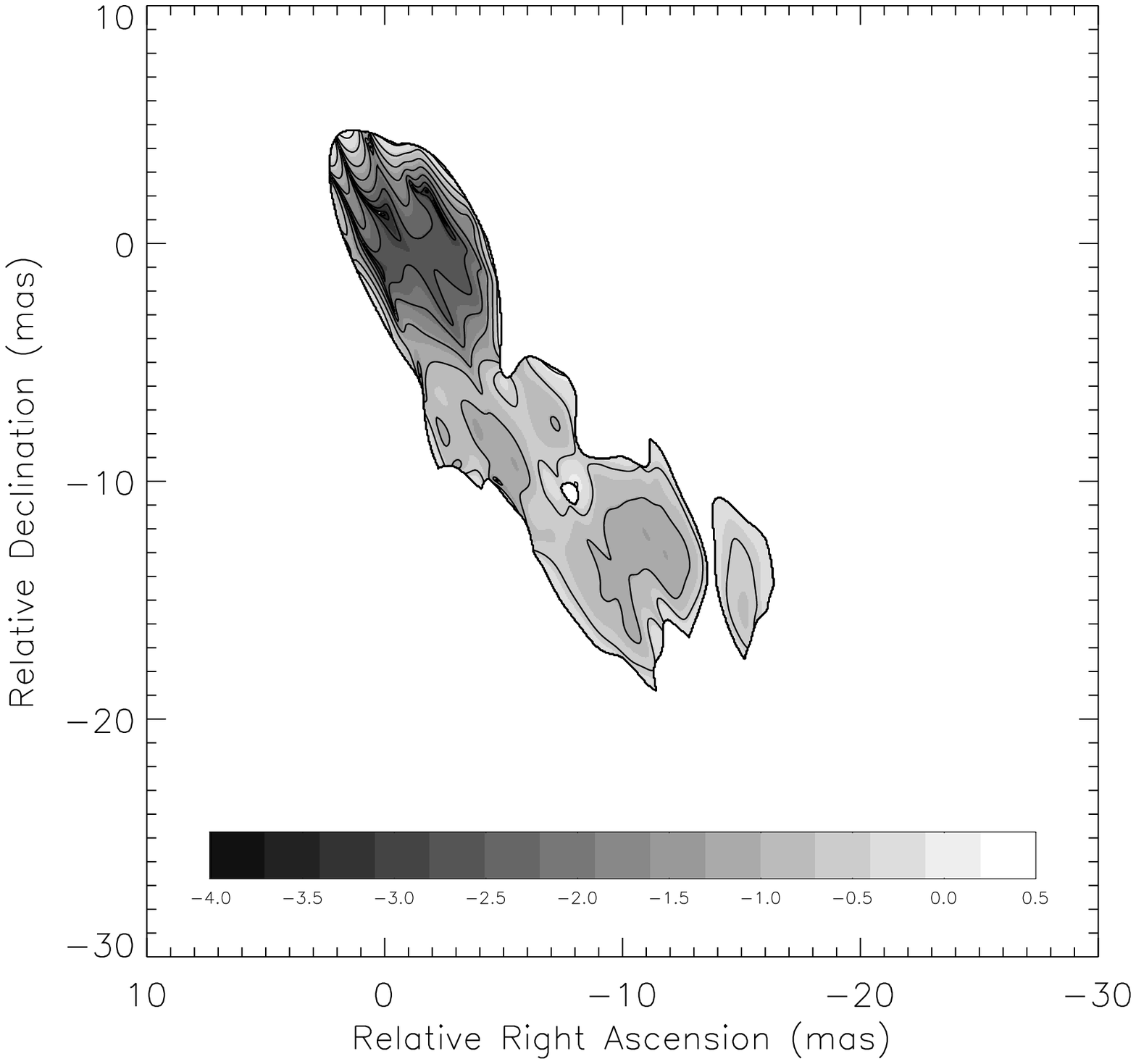}{4.0 in}{0}{60}{60}{-162}{-72}
\caption{Error map for the spectral index map presented in Figure~5.
The error was calculated by standard propagation of errors, using the fluxes at each pixel
in each residual map as the flux errors.  The gray-scale color bar along the bottom of the image
indicates the value of the logarithm of the error in the spectral index, with lighter colors
indicating a larger error.  Contours of the logarithm of the error are plotted at intervals
of 0.5, from $-3.0$ to 0.0.}
\end{figure*}

The core of 3C~279 has an inverted spectrum with steep spectral index gradients.  
Such inverted spectra are commonly interpreted as being due to self-absorption of the radio
synchrotron emission.  The calculated spectral index in the core region
ranges from $\sim$~1.0 at the western edge to the theoretical limiting value 
for synchrotron self-absorption of 2.5 
(assuming a constant magnetic field)
over a small region at the eastern edge.
Spectral indices approaching this theoretical value are almost never seen; the flatter
spectra usually observed 
are commonly interpreted as being due to an inhomogeneous source made up of a number of
synchrotron components with differing turnover frequencies (e.g. Cotton et al.\ 1980).
The highly inverted spectrum at the eastern edge may imply the detection of a homogeneous
compact component in this region, which should be an efficient producer of inverse Compton  
gamma-rays.
The spectral index gradients in the core imply
that components will have different measured separations at different frequencies.
We do indeed observe this frequency-dependent separation, the measured separation between the core and C4
model components is 3.3~mas at 5~GHz and only 2.8~mas at 1.6~GHz.  
The fact that the apparent position of the core is a function of wavelength is an important verification
of the twin exhaust model, which argues that the observed core is that position in the throat
of a nozzle where the opacity is of the order of unity.

The jet component C4 has a flat spectrum
with $\alpha$ approximately 0.25.  This is unusual, as jet components
usually have steeper spectra ($\alpha<0$).  The southwestern jet also has structure
in its spectral index distribution, with the edges of the jet appearing to have a steeper
spectrum than the center.  This spectral index structure is related to the apparent limb brightening
at 1.6~GHz which, as noted above, should be interpreted with care.
AO2 VSOP observations of 3C~279 at 1.6~GHz 
in which the baselines to the orbiting antenna have a different orientation in the $(u,v)$ plane
will allow a consistency check on the brightness and spectral index structures transverse to the jet.

\subsection{High Brightness Temperatures}
\label{tb}
Space VLBI observations have a major advantage over 
ground-based observations
because they are able to measure higher brightness temperatures.  This is
because the smallest measurable major and minor axes in equation~[\ref{tbeq}]
are proportional to the resolution of the interferometer, which is proportional to 
1/(baseline $\times$ frequency), so the denominator in equation~[\ref{tbeq}]
depends only on baseline length.
The brightness temperature limit for ground-based VLBI is $\sim 10^{11} S(1+z)/f^{2}$
and that for space VLBI with HALCA is $\sim 10^{12} S(1+z)/f^{2}$, where $S$ is the
flux density in Janskys, $z$ is the redshift, and $f$ is the smallest size that can be measured expressed
as a percentage of the beam size.
The improvement gained by space VLBI thus covers the interesting transition region around $10^{12}$~K,
the nominal inverse Compton brightness temperature limit (Kellermann \& Pauliny-Toth 1969).

Observed brightness temperatures are often used to calculate Doppler beaming factors by assuming an
intrinsic brightness temperature and using the fact that $T_{B,obs} = \delta T_{B,int}$, where
$T_{B,obs}$ is the observed source frame brightness temperature, $T_{B,int}$ is the intrinsic brightness temperature,
and $\delta$ is the Doppler factor.  The intrinsic brightness temperature depends on the physical
mechanism imposing the limitation.  Kellermann \& Pauliny-Toth (1969) showed that inverse Compton losses
limit the intrinsic brightness temperatures to $\sim~5~\times~10^{11} - 1~\times~10^{12}$~K, otherwise
the source can radiate away most of its energy on a timescale of days.  Readhead (1994) proposed that
the limiting mechanism is equipartition of energy between the particles and magnetic field, and that
intrinsic brightness temperatures are limited to $\sim~5\times~10^{10} - 1\times~10^{11}$~K.  

To be complete, brightness temperature measurements must be presented with associated errors.
Since the brightness temperature depends on the product of the major and minor axes
of the model-fit component, and these 
axis sizes can have large errors, the measured brightness temperature can also have large errors.
Error analysis of VLBI model-fit parameters has historically been problematic.  In this paper we use
the ``Difwrap'' program\footnote{http://halca.vsop.isas.ac.jp./survey/difwrap/} 
(Lovell 2000) to analyze the upper and lower limits on our
brightness temperature measurements.  This program uses the method described by Tzioumis et al.\ (1989)
in which the parameter of interest is varied in steps around the best-fit value, allowing the other
parameters to relax at each step, and the resultant model is then visually compared with the data to
determine whether or not the fit remains acceptable.  The brightness temperature of a component depends
on the flux density and size of the component, and the size depends on three of the Difmap model-fit parameters:
the major axis length, the axial ratio, and indirectly on the position angle of the major axis (since
different major axis lengths and axial ratios may be allowed at different position angles).  The size
error analysis therefore searches a three-dimensional cube in parameter space, varying the major axis
length and position angle and the axial ratio over all possible combinations given input search ranges
and step sizes.  A visual inspection is done to determine the goodness of the fit
instead of using a numerical cutoff in the chi-squared because the true number of degrees of freedom is not
well known.  Using the actual number of measured visibilities 
($\sim 10^{5}$ for these observations) to determine the degrees of freedom gives
errors that are unrealistically small, and methods used by other authors did not have a clear
physical motivation (e.g. one degree of freedom per antenna per hour [Biretta, Moore, \& Cohen 1986]).

In Table~\ref{tbtab} we show our brightness temperature error analysis for the six components in
Table~\ref{mfittab} that have best-fit brightness temperatures over $10^{12}$~K.  For each of these components
we searched a $7\times7\times7$ cube in major axis length, axial ratio, and major axis position angle.
Initially we searched major axis lengths from zero to twice the best-fit length, axial ratios from zero to
one, and a range of $\pm90\arcdeg$ in position angle; and then refined the search to a smaller grid if
necessary.  The parameter values yielding the maximum and minimum area that still gave an acceptable fit to
the data were recorded.  A similar error analysis was done for the flux density, and the extreme allowed
values of area and flux density were used to determine the maximum and minimum brightness temperatures.
Since errors in flux density and size are searched for separately, the flux density was held constant during
the size error analysis and vice-versa.  The position of the component was also held constant to avoid it
`trading identities' with another model component.  All other model components were allowed to vary.

\begin{table*}[!t]
\caption{Brightness Temperature Limits for Components with
Best-fit Brightness Temperatures $> 10^{12}$ K}
\label{tbtab}
\begin{center}
\begin{tabular}{c c r r r r r r r r r} \tableline \tableline
& & & \multicolumn{1}{c}{Min.} & \multicolumn{1}{c}{Max.\tablenotemark{b}} & & \multicolumn{1}{c}{Min.} 
& \multicolumn{1}{c}{Max.} & & \multicolumn{1}{c}{Min.} & \multicolumn{1}{c}{Max.} \\ 
Freq. & & \multicolumn{1}{c}{Flux} & \multicolumn{1}{c}{Flux} & \multicolumn{1}{c}{Flux} 
& \multicolumn{1}{c}{Area} & \multicolumn{1}{c}{Area} & \multicolumn{1}{c}{Area} 
& \multicolumn{1}{c}{$T_{B}$} & \multicolumn{1}{c}{$T_{B}$} & \multicolumn{1}{c}{$T_{B}$} \\  
(GHz) & Comp.\tablenotemark{a} & \multicolumn{1}{c}{(Jy)} & \multicolumn{1}{c}{(Jy)} & \multicolumn{1}{c}{(Jy)} 
& \multicolumn{1}{c}{(mas$^{2}$)} & \multicolumn{1}{c}{(mas$^{2}$)} & \multicolumn{1}{c}{(mas$^{2}$)} 
& \multicolumn{1}{c}{($10^{12}$ K)} & \multicolumn{1}{c}{($10^{12}$ K)} & \multicolumn{1}{c}{($10^{12}$ K)} \\ \tableline
4.8 & 1 & 1.71 & 1.61 &  ... &   0.0 &   0.0 & 0.114 &  ... & 1.15 &  ... \\ 
    & 2 & 3.91 & 3.61 &  ... & 0.110 &   0.0 & 0.146 & 2.90 & 2.01 &  ... \\ 
    & 5 & 3.39 & 3.06 & 3.60 & 0.105 & 0.070 & 0.274 & 2.63 & 0.91 & 4.19 \\ 
1.6 & 1 & 0.76 & 0.63 &  ... &   0.0 &   0.0 & 0.636 &  ... & 0.70 &  ... \\ 
    & 3 & 1.57 & 1.05 &  ... & 0.578 &   0.0 & 2.905 & 1.92 & 0.25 &  ... \\ 
    & 4 & 2.21 & 2.09 &  ... &   0.0 &   0.0 & 0.299 &  ... & 4.91 &  ... \\ \tableline
\end{tabular}
\end{center}
\tablenotetext{a}{Component numbers from the VSOP model fits in Table~\ref{mfittab}.}
\tablenotetext{b}{Upper limits to the component flux were not calculated for cases where
the component had a minimum area of zero.}
\end{table*}

Inspection of Table~\ref{tbtab} shows that
in all but one of the cases investigated the component shrinking to zero area and infinite
brightness temperature (in all cases caused by a valid fit with zero axial ratio at some position angle)
produced acceptable results,
and therefore it appears that many measured brightness temperatures, 
even those measured by space VLBI, may have error bars that extend to infinity in the positive direction.  
The measured lower limits also indicate a
considerable error in the best-fit brightness temperature values: three of the six components with best-fit
brightness temperatures over $10^{12}$~K have minimum brightness temperatures under this value.  
The other three
components have minimum brightness temperatures over $10^{12}$~K, with minimum brightness temperatures of
1.2, 2.0, and 4.9~$\times~10^{12}$~K being measured for components~1 and 2 of the 5~GHz VSOP model fit (the 
core and first jet component) and component~4 of the 1.6~GHz VSOP model fit (C4) respectively.
Bower \& Backer (1998) and Shen et al.\ (1999) report brightness temperatures of $\sim 3~\times~10^{12}$~K
from VSOP observations of NRAO~530 and PKS\,1921$-$293 respectively, but without accompanying error analyses.

If a brightness distribution other than a Gaussian is used in the model fitting,
the derived values of the brightness temperature will be different.  For example, the brightness
temperature of a homogeneous optically thick component is given by
\[T_{B}=1.77\times10^{12}\;\frac{S(1+z)}{ab\nu^{2}}\;\rm{K},\]
where the constant in front is different from that in equation~(\ref{tbeq}), and $a$ and $b$
are the lengths of the major and minor axes respectively rather than the FWHMs.
The visibility of a homogeneous optically thick component drops to 50\% at the same baseline
length as a Gaussian when its diameter equals 1.6 times the Gaussian's FWHM
(Pearson 1995), so we expect the homogeneous optically thick brightness temperature to
be about 0.6 of the Gaussian brightness temperature
(see also Hirabayashi et al.\ 1998 and in particular the correction in the erratum to this paper).
We have fit homogeneous optically thick components to the data,
and for the two components in Table~\ref{tbtab} where neither component type goes to zero size
(components~2 and 5 of the 5~GHz model fit) we measure brightness temperatures of
1.9 and 1.5~$\times~10^{12}$~K respectively for homogeneous optically thick components
rather than 2.9 and 2.6~$\times~10^{12}$~K for Gaussian components,
so we see about the expected decrease.  Since the true brightness distribution is not known,
and Gaussian components are the standard for VLBI model fitting and provide a somewhat better fit
for these observations, we remained with Gaussian components.

Our highest brightness temperature lower limit of $\sim~5~\times~10^{12}$~K for component
C4 at 1.6~GHz implies Doppler factor lower limits of 5 and 50 for the inverse Compton and
equipartition brightness temperature limits respectively.
A Doppler factor of 50 is at
the upper end of the Doppler factor distributions expected for flux-limited samples of flat-spectrum radio sources
(Lister \& Marscher 1997) and gamma-ray sources (Lister 1998).
Bower \& Backer (1998) found similar values for the Doppler factor
of NRAO~530 under these same two limiting conditions. 
If VSOP observations reveal a brightness temperature much higher than $5~\times~10^{12}$~K, or 
many brightness temperatures around $5~\times~10^{12}$~K, it may
be difficult to reconcile the high Doppler factors implied by the equipartition brightness temperature
limit with beaming statistics and with the relatively slow speeds
measured in studies of apparent velocity distributions (e.g. Vermeulen 1995).

Using an estimated speed for C4
of 7~$c$ (see $\S$~\ref{comps}), and assuming the pattern speed observed with VLBI
equals the bulk fluid speed in the jet,
bulk Lorentz factors and angles to the line-of-sight can be calculated for the jet.
For $\delta=5$, $\Gamma$=7.5 and $\theta$=11$\arcdeg$; for $\delta=50$, $\Gamma$=25.5 and $\theta$=0.3$\arcdeg$.
Again, the equipartition brightness temperature limit implies values for 3C~279 near
the extremes of expected EGRET source properties (Lister 1998).
3C~279 and NRAO~530 have both been detected by EGRET, and Bower \& Backer (1998) speculate that
blazars detected by EGRET may be those where the equipartition brightness temperature
limit is briefly (on a timescale of years) superseded by the inverse Compton catastrophe limit.
The observations presented in this paper occurred at the beginning of a total flux density flare at 5 GHz
recorded by the Michigan monitoring program that would later reach the highest flux density yet recorded 
in this program for 3C~279 at 5 GHz.
Measurements of the variability brightness temperature of this flare (L\"{a}hteenm\"{a}ki, Valtaoja, \& Wiik 1999)
together with VSOP brightness temperatures measured during AO2 should allow calculation of the intrinsic
brightness temperature and Doppler factor and allow us to estimate any departures from equipartition in this
source.

\section{Conclusions}
We have presented the first space VLBI images of 3C 279, which are the highest resolution images yet
obtained of this source at 5 and 1.6 GHz.  The parsec-scale emission is dominated by the core
and the jet component C4 which has been visible in VLBI images since 1985.
The 1.6 GHz VSOP image and the 5 and 1.6 GHz VLBA-only images
show emission from a jet extending to the southwest.  The 1.6 GHz VLBA-only image has structure that
matches that seen in the highest resolution VLA images, connecting the parsec and kiloparsec scale
structures in this source.  

We have exploited two of the main strengths of VSOP: the ability to obtain matched-resolution
images to ground-based images at higher frequencies and the ability to measure
high brightness temperatures.  The spectral index map constructed from the 1.6 GHz VSOP image and
the 5 GHz VLBA-only image has an unusually inverted spectral index in the core region, approaching
the limiting value for synchrotron self-absorption of $+2.5$.  
An extensive error analysis conducted on
the model-fit brightness temperatures reveals brightness temperature 
lower limits as high as $5~\times~10^{12}$~K.
This lower limit is significantly above both the nominal inverse
Compton and equipartition brightness temperature limits.  The derived Doppler factor, Lorentz factor, and
angle to the line-of-sight in the case of the equipartition limit
are at the upper end of the range of expected values for EGRET blazars.

\acknowledgements
Part of the work described in this paper has been carried out at the Jet
Propulsion Laboratory, California Institute of Technology, under
contract with the National Aeronautics and Space Administration.
A.E.W. acknowledges support from the NASA Long Term Space Astrophysics Program.
We gratefully acknowledge the VSOP Project, which is led by the Japanese Institute of Space and
Astronautical Science in cooperation with many organizations and radio telescopes around the world.
The National Radio Astronomy Observatory is a facility of the National Science Foundation operated
under cooperative agreement by Associated Universities, Inc.
This research has made use of 
data from the University of Michigan Radio Astronomy Observatory which is supported by
the National Science Foundation and by funds from the University of Michigan,
and the NASA/IPAC extragalactic database (NED)
which is operated by the Jet Propulsion Laboratory, California Institute of Technology, under contract
with the National Aeronautics and Space Administration.

\end{document}